# ASSESSING MULTIVARIATE PREDICTORS OF FINANCIAL MARKET MOVEMENTS: A LATENT FACTOR FRAMEWORK FOR ORDINAL DATA

By Philippe Huber,[1] Olivier Scaillet[2] and Maria-Pia Victoria-Feser[3]

*HEC–University of Geneva and Swiss Finance Institute*

Much of the trading activity in Equity markets is directed to brokerage houses. In exchange they provide so-called "soft dollars," which basically are amounts spent in "research" for identifying profitable trading opportunities. Soft dollars represent about USD 1 out of every USD 10 paid in commissions. Obviously they are costly, and it is interesting for an institutional investor to determine whether soft dollar inputs are worth being used (and indirectly paid for) or not, from a statistical point of view. To address this question, we develop association measures between what broker–dealers predict and what markets realize. Our data are ordinal predictions by two broker–dealers and realized values on several markets, on the same ordinal scale. We develop a structural equation model with latent variables in an ordinal setting which allows us to test broker–dealer predictive ability of financial market movements. We use a multivariate logit model in a latent factor framework, develop a tractable estimator based on a Laplace approximation, and show its consistency and asymptotic normality. Monte Carlo experiments reveal that both the estimation method and the testing procedure perform well in small samples. The method is then used to analyze our dataset.

**1. Introduction.** The point of departure of the present paper is the analysis of a dataset provided by the Geneva University pension fund, consisting of historical data of financial forecasts from 2 broker–dealers about the midterm evolution of the stock markets in 5 countries and the bond markets in

Received September 2007; revised May 2008.
[1]Supported in part by Swiss National Science Foundation Grant 610-057883.99. Currently at Cinetics Asset Management Geneva.
[2]Supported in part by the Swiss NCCR Finrisk and the Swiss Finance Institute.
[3]Supported in part by Swiss National Science Foundation Grant PP001-106465, and by the Swiss NCCR Finrisk.
*Key words and phrases.* Latent variable, generalized linear model, factor analysis, multinomial logit, forecasts, LAMLE, Laplace approximation.

4 zones, respectively. These broker–dealers were asked each quarter during 6 years to provide their forecasts for each country in terms of market trends (stock and bond indices) for the next 6 months. For our purpose they have been recorded on an ordinal scale from 1 to 5. In order to decide whether the forecasts are valid, they should be compared with the actual evolutions of the corresponding markets which were also recorded on the same ordinal scale. The issue is to determine whether the forecasts made by the broker–dealers are in some sense "near" the realized market evolutions six months later. Implicitly we assume that the broker–dealers are small enough so that they cannot influence the market. Therefore, there is no causal relationship between forecasts and future realizations. We are in a *multivariate* context since the forecasts concern different countries at the same time. Formally, the aim is to measure (and test for) the association between two random vectors, say, $\mathbf{X}$ (the forecasts) and $\mathbf{Y}$ (the market realizations), whose size $p \geq 2$ is the same (4 zones or 5 countries), and whose entries consist of ordinal variables corresponding to the forecast and realized market states (values in $\{1,\ldots,5\}$) for each country, respectively.

The study of the association between two random vectors is often of interest in applied statistics and econometrics. If the multivariate data are normal, the canonical correlation coefficient can be used [see, e.g., Mardia, Kent and Bibby (1979)]. It is defined as the maximal correlation coefficient between any linear combinations of elements of $\mathbf{X}$ and any linear combinations of elements of $\mathbf{Y}$. When not, for example, when the data are collected via questionnaires on scales with a limited number of points, the canonical correlation is no longer appropriate. To our knowledge, no association measure has been proposed so far to compare multivariate ordinal random variables. Association measures between univariate categorical or ordinal variables have been proposed for a long time now; see, for example, Goodman and Kruskal (1979) and Agresti (1990). For example, the Pearson tetrachoric correlation is based on the idea that there exist continuous bivariate normal distributions underlying cross-classification tables. The tetrachoric correlation is the correlation of the bivariate normal distribution having produced the cell probabilities of the table. This idea has been extended to association measures between two ordinal variables with the polychoric correlation [Olsson (1979)] and between a normal and a binary, polytomous or ordinal variable with the polyserial correlation [see Tate (1955a, 1955b), Cox (1974) and Lee and Poon (1986)].

In this paper we attempt to combine both ideas by constructing hidden or latent bivariate normal variables, one for each vector of ordinal variables and by defining an association measure which is the correlation between the latent normal variables. More precisely, we achieve that through the specification of a multivariate multinomial logit (MNL) with latent factors [see, e.g., McFadden (1984) for an introduction]. This is done in the spirit of structural equation models (SEM) with latent variables [see, e.g., Aigner et al. (1984) for an introduction] and generalized linear latent variable models [see, e.g., Bartholomew and Knott (1999) and Skrondal and Rabe-Hesketh (2004) for an introduction]. The resulting association measure is similar to the canonical correlation coefficient in the normal case and similar to the polychoric correlation in the univariate case. Our association measure corresponds to a model parameter which is easily estimated (together with other parameters) using a Laplace approximated maximum likelihood estimator (LAMLE) proposed by Huber, Ronchetti and Victoria-Feser (2004) for which we develop asymptotic properties. Consequently, statistical inference for the association measure can be performed using the properties of the estimator.

In a broader sense, latent variable models encompass a large number of models that are frequently used in recent applications. Examples include multilevel models (or hierarchical models), generalized linear mixed models or Bayesian hierarchical models. For example, Zaslavsky (2007) uses a hierarchical model for the analysis of consumer assessments of health care data, while Fielding and Yang (2005) use a generalized linear mixed model for the analysis of educational achievement data. In these models, latent variables are used to model the variability imputed to the observations that lie at different levels of clustering (i.e., the random effects) and the emphasis is put on the estimation of the variance of the random effects. Latent variables are also used to model (time) sequences by means of hidden Markov model (HMM), as is done, for example, in Zhou and Wong (2007) for the analysis of genomic sequences for short sequence elements. Wu et al. (2007) use a hierarchical state space model coupled with an HMM to analyze a short time course microarray experiment. In these models, the latent variable is the hidden time sequence modeled using a (hidden) Markov chain. Mixture models use categorical latent variables (or latent classes) mainly for classification purposes. For example, Erosheva, Fienberg and Joutard (2007) propose a latent class model to classify elderly Americans into functional disability classes according to their scores (able or not able) on different daily activities over different periods. Hence, an estimated latent score (here a class) is attributed to each observation according to the value of the response vector that permits the classification into the different latent classes. In the model we propose, the aim is also to attribute a latent score but on a continuous scale to each ordinal vector of forecasts and actual market realizations simultaneously and quantify their correlation.

Estimation of latent variable models has also seen a substantial activity in recent research. Latent variables in these models need to be integrated out from the likelihood function which then implies the computation of complicated integrals. Bayesian methods like Markov Chain Monte Carlo (MCMC) possibly coupled with the EM algorithm, are often used in practice. Alternatively, the integrals can be approximated using (adaptive) Gauss quadratures [as implemented, e.g., in GLLAMM in the STATA package; see Rabe-Hesketh, Skrondal and Pickles (2002)] or Laplace approximation. For the type of model we consider here, Huber, Ronchetti and Victoria-Feser (2004) argue that a Laplace approximation of the likelihood function is a better approach: the resulting estimator is asymptotically unbiased and fast to compute (a key advantage when using resampling methods, for example). Finally, it should be noted that standard SEM packages rely on two-stage procedures that basically reduce the information given in the sample to an estimate of the (multivariate) mean and covariance (using, e.g., polychoric correlations). These two-stage procedures are slower and cannot guarantee consistency of parameter estimators.

The paper is organized as follows. In Section 2 we present the datasets and the problem at hand, and motivate the construction of an association measure between the ordinal random vectors. In Section 3 we develop a multivariate multinomial logit (MNL) with latent factors in the framework of the generalized linear latent variable model (GLLVM) and propose the correlation between the latent variables as the measure of association. We then compare in Section 4 this measure to the polychoric correlation and the canonical correlation. Estimation and asymptotic properties are investigated in Section 5. We rely on the so-called Laplace approximation [De Bruijn (1981)] to get a tractable and fast estimation procedure of the latent variable model, and show consistency and asymptotic normality of the resulting estimators. Section 6 is devoted to Monte Carlo experiments aimed at gauging the performance in small samples of the estimation method and the testing procedure of the measure of association. We gather the empirical results in Section 7, while technical details and proofs are relegated to supplemental material [see Huber, Scaillet and Victoria-Feser (2009c)].

**2. The data.** The database[4] contains the forecasts (in terms of trends) of two broker–dealers $A$ and $B$ about the mid-term (6 months) evolution of the stock market in five different countries (Switzerland, Germany, France, Great Britain and USA) for $A$ and the bond market in four zones (Switzerland, Euro Zone, Great Britain and USA) for $B$. The trends have been clearly and precisely defined as corresponding to a given future variation $x$ with: $x < -10\%$ (strong bear), $-10\% < x < -5\%$ (bear), $-5\% < x < 5\%$ (neutral), $5\% < x < 10\%$ (bull), $10\% < x$ (strong bull), for the stock market and $x < -0.25\%$, $-0.25\% < x < -0.10\%$, $-0.10\% < x < 0.10\%$, $0.10\% < x < 0.25\%$, $0.25\% < x$, for the bond market. They have been recorded on an ordinal scale from 1 to 5. In both cases, we compare the forecasts to the actual returns of the corresponding markets six months later. For the stock market, the actual trends are measured on the stock indices: S&P500 (US), FTSE100 (UK), CAC40 (FR), DAX (D) and SMI (CH). The sample starts in July 1997 and finishes in April 2003 with one forecast every quarter (22 observations). The observations are sequential in time but since the time gaps are of 3 months (a quarter), we can assume that there is no serial correlation.

---

[4]The datasets are provided as part of supplemental material; see Huber, Scaillet and Victoria-Feser (2009a).

The data corresponding to broker–dealer $A$ (stock markets) are represented in Figure 1 in the form of graphs of the observed versus predicted values for each stock market separately. The data corresponding to broker–dealer $B$ (bond markets) are represented in the same way in Figure 2. It is not clear at all if the predictions are "in general" in accordance with the corresponding actual market values. It seems, however, that roughly the broker–dealers are in general able to follow the trends, except that they tend to underestimate their magnitude.

This type of representation gives a first idea on the performance of the broker–dealers, but a more formal approach, in the form of an indicator (and its variability), is more appropriate. We need to compare the predictive performance of several broker–dealers for these markets from an institutional point of view. If the data were recorded on a normal scale, a canonical correlation could be used. For an ordinal scale, an appropriate measure of correlation is the polychoric correlation, but it is only defined between pairs of (ordinal) variables. Hence, at least for the problem at hand, we need an association measure between two vectors of ordinal variables.

Measuring the predictive ability of broker–dealers is an important issue because institutional investors make a large portion of overall trading volume in Equity markets, and much of this trading activity is directed to brokerage houses who execute trades. In exchange for directed trades, most of the brokerage houses provide so-called "soft dollars." Soft dollar arrangements are arrangements under which products or services other than execution of securities transactions are obtained by an institutional investor from or through a broker–dealer in exchange for the placement of his orders [see Blume (1993), Johnsen (1994) and the Securities and Exchange Commission (1998) for the detailed definition, history and law related to soft dollars]. These arrangements are best thought of as ways of subsidizing the research inputs that investors use to identify profitable trading opportunities. In contrast to "hard dollars" (actual cash), which have to be reported on investor books, soft dollars are incorporated into brokerage fees and the expenses investors pay for needs not be reported directly. Soft dollars arrangements were first developed as a means of competition among brokers. With broker–dealers being unable to compete based on commission rates fixed by regulation, complex arrangements to provide equity research and services became a primary tool of differentiation among brokerage houses. They also provided a way to investors to recapture a portion of the high commissions they were required to pay. The US Security Exchange Commission (SEC) abolished fixed rate commissions on May 1, 1975. Shortly after industry participants expressed concern that soft dollar practices would be viewed as a violation of a manager fiduciary obligation to place the client interest above his own. In response, US Congress passed Section 28(e) of the Securities and Exchange Act to provide a "safe harbor" and protect managers of being accused of breaching their duty. This legal acceptance explains why the industry still offers such arrangements nowadays. US regular surveys about the size of the soft dollar industry are conducted by Greenwich Associates. Their 2007 survey of 229 financial institutions indicates that soft dollar commissions totaled almost USD 723 million in 2007 down from USD 970 million in 2006. This repre-sents about USD 1 out of every USD 10 (10%!) paid in commissions by those firms involved. As recently as 2004, more than 80% of institutions used soft dollars; by 2007 that proportion is still around 60%. On the contrary, less than a third are currently buying equity research and services with hard dollars. Obviously soft dollar practices are costly and widespread, and it is therefore important for an institutional investor to determine whether these soft dollar inputs are worth being used (and indirectly paid for) or not, from a statistical point of view. Indeed, even if soft dollars vary by account size, annual turnover and asset class, they still represent a meaningful portion of the overall annual cost of actively managed equity portfolios. They directly impact the overall performance of funds, and thus are to be monitored.

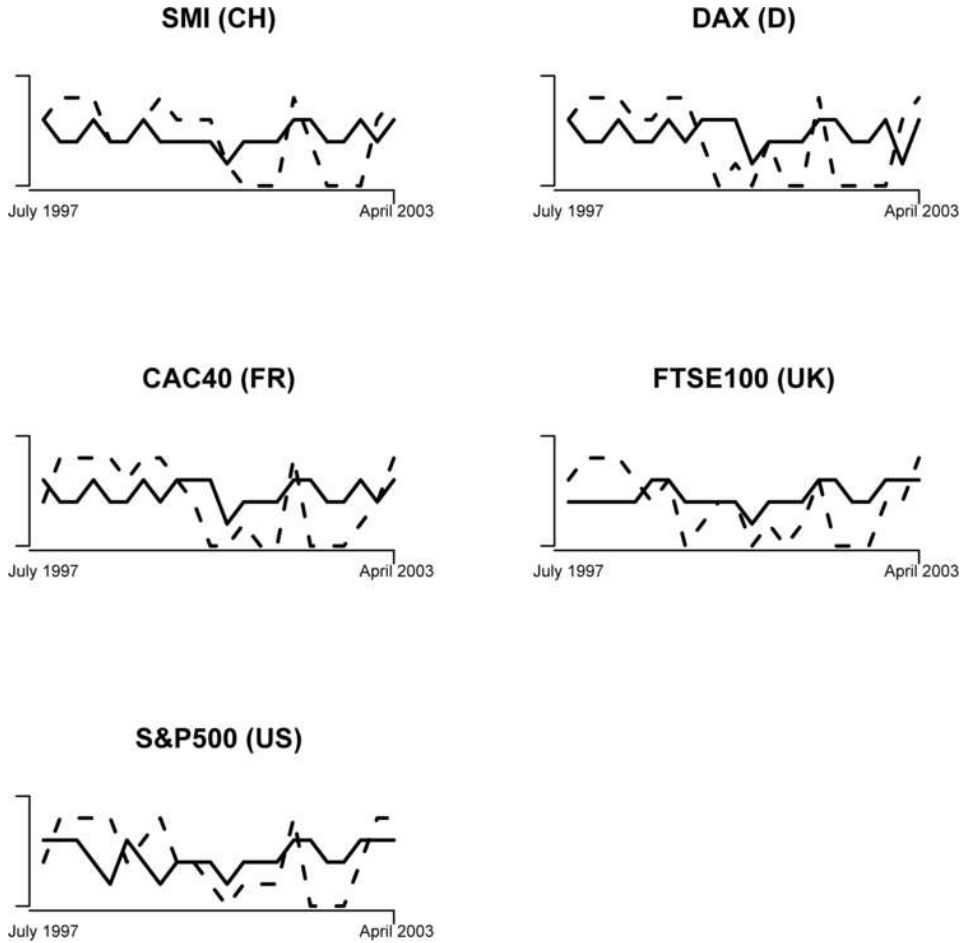

Fig. 1. *Observed (dashed line) and predicted (solid line) market values by broker–dealer A, for five different stock markets.*

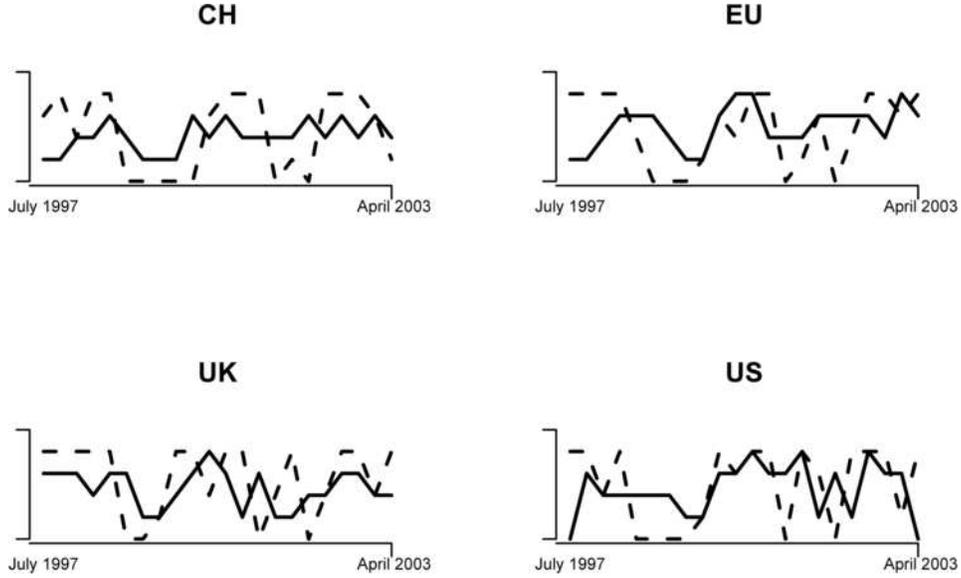

Fig. 2. *Observed (dashed line) and predicted (solid line) market values by broker–dealer B, for four different zones.*

After presenting the model in Section 3, we will assess in Section 7 the predictive ability of the broker–dealers by means of our association measure.

**3. A SEM for multivariate measure of association.** Recall that the aim is here to measure an association between a set $\mathbf{X} = (X^{(1)}, \ldots, X^{(p)})'$ of manifest variables and another set $\mathbf{Y} = (Y^{(1)}, \ldots, Y^{(p)})'$ of manifest variables. Let $\mathbf{U} = (\mathbf{X}', \mathbf{Y}')'$ and let $\mathbf{F}_X$ and $\mathbf{F}_Y$ be latent variable vectors of dimension $m_X \times 1$ and $m_Y \times 1$ with $m_Y, m_X < p$. Let also $\mathbf{F} = (1, \mathbf{F}'_X, \mathbf{F}'_Y)' = (1, \mathbf{F}'_{(2)})'$, such that $\mathbf{F}_{(2)} \sim N(\mathbf{0}, \mathbf{R})$, $\mathbf{R}$ being a correlation matrix. If the manifest variables are normal, we could use a SEM for normal vectors given by

(1) $$U^{(l)} | \mathbf{F} \overset{\text{ind.}}{\sim} N(\boldsymbol{\lambda}'_l \mathbf{F}, \phi_l^2), \qquad l = 1, \ldots, 2p$$

with, for $l = 1, \ldots, p$, $\boldsymbol{\lambda}_l = (\alpha_X^{(l)}, \boldsymbol{\beta}_X^{(l)\prime}, \mathbf{0}')'$ the intercept and factor loadings for the manifest variables $\mathbf{X}$ and $\phi_l^2$ their residual variance, and, for $l = p+1, \ldots, 2p$, $\boldsymbol{\lambda}_l = (\alpha_Y^{(l-p)}, \mathbf{0}', \boldsymbol{\beta}_Y^{(l-p)\prime})'$ the intercept and factor loadings for the manifest variables $\mathbf{Y}$ and $\phi_l^2$ their residual variance. In principle, the dimension of the latent spaces $m_Y, m_X$ could be greater than 1, but for the purpose of building a single association measure, the choice is $m_Y = m_X = 1$. Consequently, $\mathbf{R} = \begin{bmatrix} 1 & \rho \\ \rho & 1 \end{bmatrix}$ has a single unknown parameter $\rho$ which can be interpreted as an association measure between $\mathbf{X}$ and $\mathbf{Y}$ (see also Section 4).

The normalization (unit variance) of the latent variable $\mathbf{F}_{(2)}$ is necessary for identification purposes. Indeed, since the latent variables in model (1) are multiplied by the factor loadings, an increase (decrease) of the variance of $\mathbf{F}_{(2)}$ cannot be distinguished from a simultaneous decrease (increase) of the factor loadings. Moreover, since we are interested in computing the correlation between the latent factors $\mathbf{F}_X$ and $\mathbf{F}_Y$, this normalization has no influence on the association measure.

Incorporating additional factors (i.e., increasing $m_Y$ and $m_X$) does not raise any difficulties; however, the interpretation in terms of measure of association becomes more difficult. It is therefore important to check the model fit when assuming $m_X = m_Y = 1$; this will be done in our empirical illustration (see Section 7). Finally, we also note that the dimensions of $\mathbf{X}$ and $\mathbf{Y}$ need not to be equal.

Model (1) provides measures of association between vectors of normal variables. It can be generalized to the family of exponential distributions for vectors $\mathbf{Z} = (\mathbf{X}', \mathbf{Y}')'$ of (nonnecessarily normal) manifest variables with probability distribution function:

$$(2) \qquad g_l(Z^{(l)}|\mathbf{F}) = \exp\left\{\frac{u(\boldsymbol{\lambda}_l'\mathbf{F})Z^{(l)} - b(u(\boldsymbol{\lambda}_l'\mathbf{F}))}{\phi_l} + c(Z^{(l)}, \phi_l)\right\}, \qquad l = 1, \ldots, 2p,$$

where $u(\boldsymbol{\lambda}_l'\mathbf{F})$ is the so-called canonical parameter, $b(u(\boldsymbol{\lambda}_l'\mathbf{F}))$ and $c(z^{(l)}, \phi_l)$ are specific functions whose form depends on the particular exponential distribution, and $\phi_l$ is a scale parameter [see McCullagh and Nelder (1989)]. Except for the normal case, the expectation $E[Z^{(l)}|\mathbf{F}]$ is not a linear function of $\mathbf{F}$, but is linked to the linear predictor through a link function $\nu$ as

$$\nu(E[Z^{(l)}|\mathbf{F}]) = \boldsymbol{\lambda}_l'\mathbf{F}.$$

We further have that $u(\boldsymbol{\lambda}_l'\mathbf{F}) = \boldsymbol{\lambda}_l'\mathbf{F}$ when we choose the so-called canonical link function for $\nu$. This model belongs to the class of Generalized Linear Latent Variables Model (GLLVM) which has been proposed by Moustaki

(1996) and Moustaki and Knott (2000) under an assumption of independence between the Gaussian latent variables (diagonal $R$). This type of modeling can be viewed as an extension of the usual Generalized Linear Models approach [McCullagh and Nelder (1989)] to the latent factor framework.

The conditional independence of the manifest variables $Z^{(l)}$ given the latent ones is assumed, so that the conditional joint density of the manifest variables is $\prod_{l=1}^{2p} g_l(Z^{(l)}|\mathbf{F})$ and their marginal joint distribution is

$$(3) \qquad f(\mathbf{z}) = \int \left[\prod_{l=1}^{2p} g_l(Z^{(l)}|\mathbf{F})\right] \psi(\mathbf{F})\, d\mathbf{F},$$

with $\psi$ being the $N(\mathbf{0}, \mathbf{R})$ density function.

Because of the nature of the data at hand, we need to develop hereafter the case of ordinal variables, that is, ordered categorical variables. Let $Z^{(l)}|\mathbf{F}$ follow a multinomial distribution with possible values (or categories) going from 1 to $q_l$. In the following we opt for a cumulative logit formulation [see Agresti (1990) for the advantages of this formulation over other ones, and Jöreskog and Moustaki (2001) for a comparison of different approaches in the framework of factor analysis with ordinal data] to account for the ordered nature of the categorical data. Let $P_{ls} = P[Z^{(l)} \leq s|\mathbf{F}]$, $s = 1, \ldots, q_l$, be the conditional cumulative distribution functions. The quantity $\log(P_{ls}/(1 - P_{ls}))$ is the log-odds of falling into or below a category $s$ versus falling above it for the manifest variable $l$. It is used in the logit link between the linear predictor and the conditional cumulative probability distribution:

$$(4) \qquad \nu(P_{ls}) = \log\left(\frac{P_{ls}}{1 - P_{ls}}\right) = \boldsymbol{\lambda}'_{ls}\mathbf{F},$$

where $\boldsymbol{\lambda}_{ls} = (\alpha_X^{(l,s)}, \boldsymbol{\beta}_X^{(l)\prime}, \mathbf{0}')'$ or $\boldsymbol{\lambda}_{ls} = (\alpha_Y^{(l-p,s)}, \mathbf{0}', \boldsymbol{\beta}_Y^{(l-p)\prime})'$ depending on the value of $l$. The index $s$ in the $\alpha$'s indicates that each intercept depends not only on the manifest variable $l$ but also on the category $s$. The constraint for each manifest variable that the $p$ slope coefficients (the beta's) in $\boldsymbol{\lambda}_{ls}$ does not depend on the category $s$ is known as the proportional-odds assumption, and essentially allows us to reduce the number of parameters to be estimated. The intercepts take the interpretation of thresholds and are monotonic in the sense that the lowest category receives the lowest threshold, and so on. They represent the log-odds of falling into or below category $s$ when all latent variables are nil, while a given positive slope leads to an increase on the log-odds of falling into or below any category associated with a one unit increase in the corresponding latent variable. A positive slope indicates thus an increase in the odds themselves, and higher probabilities for the manifest variable to take low values. For identification purposes, the highest threshold is set equal to infinity by convention, which means that we only need to estimate the $q_l - 1$ threshold for the manifest variable $l$. With all these restrictions, the model is fully identifiable. In general, the thresholds can be assumed to be different for each manifest (ordinal) variable. However, when the ordinal variables take values under the same measurement unit (e.g., percentages), we can constrain the thresholds to be equal for all manifest variables, that is, $\alpha_X^{(l,s)} = \alpha_Y^{(l-p,s)} = \alpha^{(s)}$ for all $l$. This is a suitable constraint for the analysis of our data (see Section 7).

The scale parameter $\phi_l$ is here equal to 1, while the canonical parameter is not linear in the latent factors (since we do not use the canonical link function), but equal to

$$u(\boldsymbol{\lambda}_{ls}'\mathbf{F}) = \log\left(\frac{P_{ls}}{P_{l,s+1} - P_{ls}}\right),$$

while

$$b(u(\boldsymbol{\lambda}_{ls}'\mathbf{F})) = \log(1 + \exp(u(\boldsymbol{\lambda}_{ls}'\mathbf{F})))$$
$$= \log\left(\frac{P_{l,s+1}}{P_{l,s+1} - P_{ls}}\right),$$

and $c(Z^{(l)}, \phi_l) = 0$. The conditional distribution of the manifest variable is given by

$$g(Z^{(l)}|\mathbf{F}) = \prod_{s=1}^{q_l}(P_{ls} - P_{l,s-1})^{\iota(Z^{(l)}=s)}$$

(5)
$$= \prod_{s=1}^{q_l-1}\left(\frac{P_{ls}}{P_{l,s+1}}\right)^{\iota(Z^{(l)}\leq s)}\left(\frac{P_{l,s+1} - P_{ls}}{P_{l,s+1}}\right)^{\iota(Z^{(l)}\leq s+1)-\iota(Z^{(l)}\leq s)}$$
$$= \exp\left(\sum_{s=1}^{q_l-1}[\iota(Z^{(l)}\leq s)u(\boldsymbol{\lambda}_{ls}'\mathbf{F}) - \iota(Z^{(l)}\leq s+1)b(u(\boldsymbol{\lambda}_{ls}'\mathbf{F}))]\right),$$

where $\iota(Z^{(l)} = s) = 1$ if $Z^{(l)} = s$ and 0 otherwise, and $\iota(Z^{(l)} \leq s) = 1$ if $Z^{(l)} \leq s$, and $\iota(Z^{(l)} \leq s) = 0$ otherwise.

Note that we could use a probit link instead of the logit function. In practice, however, the difference is very small since these two link functions are very close ($|\Phi(x) - \Psi(1.7x)| < 0.01$, $\forall x$, where $\Psi$ is the logistic distribution function and $\Phi$ the normal cumulative distribution function); see, for example, Lord and Novick (1968). In the regression model (i.e., $\mathbf{F}$ is observed and $\mathbf{Z}$ is the univariate ordinal response variable), McCullagh and Nelder (1989) use the same approach to link the explanatory variable to the first moments of the response variable. Finally, let us remark that a specification in terms of latent variables is a usual way to reduce the complexity of multinomial model calculation [see McFadden (1984), page 1419], and to achieve a relative parsimony in the modeling. This is even more relevant, if not inevitable, in a multivariate framework.

**4. Relationship with other measures of association.** In this section, we compare the measure of association $\rho$ with two other measures, namely, the polychoric correlation which associates two ordinal random vectors and the canonical correlation which associates two vectors of normal variables.

There are many other association measures between two ordinal variables such as Goodman and Kruskal (1954) and Gamma and Kendall (1945) tau-$b$ [see also Agresti (1984)], but the polychoric correlation is the most similar to our measure.

The polychoric correlation is based on the assumption of the existence of a vector $(X^*, Y^*)$ of bivariate normal variables with zero mean, unit variance and correlation $\rho$. Instead of observing directly $(X^*, Y^*)$, we observe the vector $(X, Y)$ of ordered multinomial variables, taking ordered values in, say, $\{1, \ldots, q_X\}$ and $\{1, \ldots, q_Y\}$, respectively. The observed variables are linked to $(X^*, Y^*)$ by means of a set of thresholds $\alpha_X^{(s_X)}, \alpha_Y^{(s_Y)}, s_x = 1, \ldots, q_X, s_Y = 1, \ldots, q_Y$ through $P(X \leq s_X, Y \leq s_Y) = P(X^* \leq \alpha_X^{(s_X)}, Y^* \leq \alpha_Y^{(s_Y)})$. Using the bivariate normal as the (indirect) model, we can estimate the so-called polychoric correlation $\rho$ by using the likelihood function given in Huber, Scaillet and Victoria-Feser (2009a) [see also Olsson (1979)]. With the GLLVM with conditional density (5) when $p = 1$ and taking the probit link instead of the logit link, we get a likelihood function that shows that although the estimator of the correlation $\rho$ is different from the polychoric correlation [see Huber, Scaillet and Victoria-Feser (2009a)], they are directly comparable. Indeed, for both estimators an assumption is made about the distribution of the underlying (or factor) variables, namely, normality, and the thresholds need to be estimated together with the correlation coefficient. With the GLLVM, an additional centering $\beta_X F_X$ ($\beta_Y F_Y$, resp.) is also needed. However, the advantage of the GLLVM approach is that it can be easily extended to multivariate $\mathbf{X}$ and $\mathbf{Y}$, which is not the case with the polychoric correlation.

On the other hand, the canonical correlation can be used to assess the association between two multivariate normal variables [see, e.g., Mardia, Kent and Bibby (1979)]. It is defined as the maximal correlation coefficient between any linear combinations of $\mathbf{X}$ and any linear combinations of $\mathbf{Y}$. For a moment suppose that $(\mathbf{X}', \mathbf{Y}')'$ is distributed as a multivariate normal random variable with mean $(\boldsymbol{\mu}'_X, \boldsymbol{\mu}'_Y)'$ and covariance matrix

$$\boldsymbol{\Sigma} = \begin{pmatrix} \boldsymbol{\Sigma}_{XX} & \boldsymbol{\Sigma}_{XY} \\ \boldsymbol{\Sigma}'_{XY} & \boldsymbol{\Sigma}_{YY} \end{pmatrix}.$$

The canonical correlation coefficient is then defined by

$$\rho_c = \frac{\mathbf{b}_X^{*\prime}\boldsymbol{\Sigma}_{XY}\mathbf{b}_Y^*}{\sqrt{\mathbf{b}_X^{*\prime}\boldsymbol{\Sigma}_{XX}\mathbf{b}_X^*\mathbf{b}_Y^{*\prime}\boldsymbol{\Sigma}_{YY}\mathbf{b}_Y^*}}, \tag{6}$$

where $\mathbf{b}_X^*$ and $\mathbf{b}_Y^*$ are the solutions to the maximization problem

$$\max_{\mathbf{b}_X,\mathbf{b}_Y} \frac{\mathbf{b}_X^\prime\boldsymbol{\Sigma}_{XY}\mathbf{b}_Y}{\sqrt{\mathbf{b}_X^\prime\boldsymbol{\Sigma}_{XX}\mathbf{b}_X\mathbf{b}_Y^\prime\boldsymbol{\Sigma}_{YY}\mathbf{b}_Y}}. \tag{7}$$

As noticed in Section 3, in the GLLVM with normal manifest variables and with $m_X = m_Y = 1$, the correlation $\rho$ between the latent variables can be interpreted as an association measure between $\mathbf{X}$ and $\mathbf{Y}$. We may therefore ask the next question: what is the link between the latter and the standard canonical correlation coefficient $\rho_c$ given in (6)? The following proposition shows that $\rho$ can be rewritten in an analogous form to (6), namely, a correlation between linear combinations of $\mathbf{X}$ and linear combinations of $\mathbf{Y}$ but with a modified covariance matrix $\boldsymbol{\Sigma}^* = \boldsymbol{\Sigma} + \boldsymbol{\psi}$ with $\boldsymbol{\psi} = \text{diag}(\phi_l^2)$.

PROPOSITION 1. *For the SEM (1), the correlation coefficient $\rho$ between the latent variables is given*

$$\rho = \frac{\boldsymbol{\beta}_X^\prime\boldsymbol{\Sigma}_{XY}^*\boldsymbol{\beta}_Y}{\sqrt{\boldsymbol{\beta}_X^\prime\boldsymbol{\Sigma}_{XX}^*\boldsymbol{\beta}_X\boldsymbol{\beta}_Y^\prime\boldsymbol{\Sigma}_{YY}^*\boldsymbol{\beta}_Y}}, \tag{8}$$

*with $\boldsymbol{\beta}_X = (\beta_X^{(1)},\ldots,\beta_X^{(p)})^\prime$, $\boldsymbol{\beta}_Y = (\beta_Y^{(1)},\ldots,\beta_Y^{(p)})^\prime$ and $\boldsymbol{\Sigma}^* = \boldsymbol{\Sigma} + \boldsymbol{\psi}$.*

Equation (8) can be interpreted as the correlation between any linear combinations of elements of $\mathbf{X}$ and any linear combinations of elements of $\mathbf{Y}$ when the covariance structure is accounted for measurement error in the manifest variables via $\boldsymbol{\psi}$. The correlation coefficient $\rho$ is thus different from the canonical correlation coefficient $\rho_c$ by construction. The advantage of defining $\rho$ instead of $\rho_c$ as an association measure between two vectors of random variables is that $\rho$ can be easily generalized to the case of nonnormal variables, like, for example, ordinal variables in our case.

Keller and Wansbeek (1983) mention that the canonical coefficients can be obtained when $\boldsymbol{\psi}$ has a particular form. They also use the SEM in (1) with categorical variables, and show the relationships between the resulting models and Correspondence Analysis. Our approach here is different since in the nonnormal case (e.g., ordinal) we change the SEM model to take into account the specificity of the data.

**5. Estimation and asymptotic properties.** In this section we propose an estimator for the association measure $\rho$ induced by the GLLVM and derive its asymptotic properties. Traditionally, the parameters of a GLLVM, when the manifest variables are ordered ordinal, are estimated by means of a two-stage approach. Polychoric correlations are estimated between all pairs of ordinal variables and used in the construction of the correlation matrix between the manifest variables [Muthén (1984), Poon and Lee (1987)]. Then a traditional factor analysis (with or without correlated latent variables) is performed. Inference on the GLLVM parameters is performed using the asymptotic properties of the polychoric correlation [see, e.g., Jöreskog (1994)]. This two-stage approach is based on the assumption that the underlying distribution of the manifest variables is normal. When this is not the case, the resulting estimators can be biased, essentially because the information in the sample is reduced to estimates of the first two moments of the multivariate distribution of the ordinal variables (i.e., their mean and correlation matrix). By means of simulations, Huber, Ronchetti and Victoria-Feser (2004) show the bias effect of this two-stage estimation procedure on estimates of GLLVM with mixtures between binary and normal variables, while Elefant-Yanni, Huber and Victoria-Feser (2004) examine this effect on estimates of GLLVM with ordinal variables.

To describe the estimation procedure, let $\mathbf{z} = [\mathbf{z}_1, \ldots, \mathbf{z}_n]'$, with $\mathbf{z}_i = [z_i^{(1)}, \ldots, z_i^{(2p)}]'$, $n$ the sample size, and $2p$ the number of manifest variables. As the marginal distribution of the observed variable must be integrated out from the conditional distributions $g(Z^{(l)}|\mathbf{F})$ given by (3), we use a Laplace approximation [see De Bruijn (1981)] to approximate the likelihood function of the sample as it has been done in Huber, Ronchetti and Victoria-Feser (2004) for other types of variables.

The Laplace approximation to integrals goes back to the original work of Laplace. This technique is widely used in mathematics; see, for example, De Bruijn (1981). In statistics, it has been used successfully to approximate posterior distributions in Bayesian statistics [see, e.g., Tierney and Kadane (1986)] and in relation to saddlepoint approximations [Field and Ronchetti (1990)].

Let $h:\mathbb{R}^m \to \mathbb{R}$ be a function which satisfies the following conditions: it is continuous and has a global maximum in $\hat{\mathbf{x}}$, its first and second derivatives exist in a neighborhood of $\hat{\mathbf{x}}$ and $\partial h(\hat{\mathbf{x}})/\partial \mathbf{x} = 0$ and $\mathbf{H}(\hat{\mathbf{x}}) = \partial^2 h(\hat{\mathbf{x}})/\partial \mathbf{x}\,\partial \mathbf{x}'$, the Hessian matrix, is such that $-\mathbf{H}(\hat{\mathbf{x}})$ is positive definite. Moreover, $h(\mathbf{x})$ is sharply peaked in the neighborhood of $\hat{\mathbf{x}}$, that is, two positive scalars $b$ and $c$ exist such that

$$h(\mathbf{x}) \leq -b \qquad \text{if } |\hat{\mathbf{x}} - \mathbf{x}| \geq c.$$

Then,

$$(9) \quad \int e^{th(\mathbf{x})}\, d\mathbf{x} = (2\pi)^{m/2} \det(-\mathbf{H}(\hat{\mathbf{x}}))^{-1/2} t^{-m/2} \exp(th(\hat{\mathbf{x}}))(1 + O(t^{-1})),$$
$$t \to \infty.$$

Equation (9) is obtained by an expansion of $h(x)$ about its maximum $\hat{\mathbf{x}}$:

$$\int e^{th(\mathbf{x})}\, d\mathbf{x} \approx \int \exp\left(th(\hat{\mathbf{x}}) + t\frac{\partial}{\partial \mathbf{x}}h(\hat{\mathbf{x}})(\mathbf{x}-\hat{\mathbf{x}}) + \frac{1}{2}t(\mathbf{x}-\hat{\mathbf{x}})\mathbf{H}(\hat{\mathbf{x}})(\mathbf{x}-\hat{\mathbf{x}})'\right) d\mathbf{x}$$
$$= \exp(th(\hat{\mathbf{x}})) \int \exp\left(\frac{1}{2}t(\mathbf{x}-\hat{\mathbf{x}})\mathbf{H}(\hat{\mathbf{x}})(\mathbf{x}-\hat{\mathbf{x}})'\right) d\mathbf{x}$$
$$= (2\pi)^{m/2} \det(-\mathbf{H}(\hat{\mathbf{x}}))^{-1/2} t^{-m/2} \exp(th(\hat{\mathbf{x}})).$$

Letting $\boldsymbol{\lambda}$ denote the vector of all loadings and thresholds, and $\mathbf{R}$ the correlation matrix of the latent factors, the approximated log-likelihood $\tilde{l}$ for a model with ordered multinomial distributed manifest variables is [see Huber, Scaillet and Victoria-Feser (2009c)]

$$\tilde{l}(\boldsymbol{\lambda}, \mathbf{R}|\mathbf{z}) = \sum_{i=1}^{N}\left(-\frac{1}{2}\log\det(\boldsymbol{\Gamma}(\boldsymbol{\lambda},\mathbf{R},\widehat{\mathbf{F}}_i)) - \frac{1}{2}\log\det(\mathbf{R})\right.$$
$$+ \sum_{l=1}^{2p}\sum_{s=1}^{q_l-1}[\iota(z_i^{(l)} \le s)u(\boldsymbol{\lambda}_{ls}'\hat{\mathbf{F}}_i)$$
$$(10)$$
$$- \iota(z_i^{(l)} \le s+1)\log(1 + \exp u(\boldsymbol{\lambda}_{ls}'\hat{\mathbf{F}}_i))]$$
$$\left. - \frac{\widehat{\mathbf{F}}_{i(2)}'\mathbf{R}^{-1}\widehat{\mathbf{F}}_{i(2)}}{2}\right),$$

where $\boldsymbol{\Gamma}(\boldsymbol{\lambda},\mathbf{R},\widehat{\mathbf{F}}_i)$ is a correction matrix that comes from the Laplace approximation, $\hat{\mathbf{F}}_{i(2)} = [\hat{\mathbf{F}}_{iX}', \hat{\mathbf{F}}_{iY}']'$ and $\hat{\mathbf{F}}_i = [1, \hat{\mathbf{F}}_{i(2)}']'$ is the estimator of the latent score for the $i$th observation which is given by the implicit equation

$$\hat{\mathbf{F}}_{i(2)} := \hat{\mathbf{F}}_{i(2)}(\boldsymbol{\lambda},\mathbf{R},\mathbf{z}_i) = \sum_{l=1}^{2p}\sum_{s=1}^{q_l-1}(\iota(z_i^{(l)} \le s)P_{l,s+1}(\boldsymbol{\lambda}_{ls}'\hat{\mathbf{F}}_i)$$
$$(11)$$
$$- \iota(z_i^{(l)} \le s+1)P_{ls}(\boldsymbol{\lambda}_{ls}'\hat{\mathbf{F}}_i))\mathbf{R}\boldsymbol{\lambda}_{l(2)},$$

where $\boldsymbol{\lambda}_{l(2)}$ is $\boldsymbol{\lambda}_{ls}$ without its first element.

Huber, Ronchetti and Victoria-Feser (2004) point out that $\hat{\mathbf{F}}_{i(2)}$ can be seen as the MLE of $\mathbf{F}_{i(2)}$. The Laplace Approximated MLE (LAMLE) of the models parameters are obtained from the optimization of $\tilde{l}$, whose derivatives can be computed explicitly, but are omitted here for sake of space. Hereafter, we establish the consistency and asymptotic normality of the LAMLE $\widehat{\boldsymbol{\theta}}$ of $\boldsymbol{\theta} = (\boldsymbol{\lambda}', \text{vech}(\mathbf{R})')'$, where $\text{vech}(\mathbf{A})$ is the stack of the elements on and below the diagonal of $\mathbf{A}$.

PROPOSITION 2 (Consistency). *Let $\boldsymbol{\theta} \in \Theta$. If $\Theta$ is compact,*

$$\widehat{\boldsymbol{\theta}} \xrightarrow{p} \boldsymbol{\theta}_0 = \arg\max_{\boldsymbol{\theta} \in \Theta} E[\widetilde{l}(\boldsymbol{\theta})].$$

Note that the empirical approximated likelihood is here too complex to be shown to be concave in $\boldsymbol{\theta}$. Under concavity of the objective function, compactness can be replaced by the assumption of $\boldsymbol{\theta}_0$ being an element of the interior of a convex set $\Theta$ [see, e.g., Theorem 2.7 of Newey and McFadden (1994)].

PROPOSITION 3 (Asymptotic normality). *If $\Theta$ is compact, $\boldsymbol{\theta}_0 \in \text{interior}(\Theta)$, and $J_0 = E[\partial^2 \widetilde{l}(\boldsymbol{\theta}_0)/\partial \boldsymbol{\theta}\, \partial \boldsymbol{\theta}']$ is nonsingular,*

$$\sqrt{T}(\hat{\boldsymbol{\theta}} - \boldsymbol{\theta}_0) \xrightarrow{d} N(0, \mathbf{J}_0^{-1} \mathbf{I}_0 \mathbf{J}_0^{-1}),$$

*with $\mathbf{I}_0 = E[\partial \widetilde{l}(\boldsymbol{\theta}_0)/\partial \boldsymbol{\theta}\, \partial \widetilde{l}(\boldsymbol{\theta}_0)/\partial \boldsymbol{\theta}']$.*

Alternative estimators have been proposed in the framework of Generalized Mixed Linear Models, such as the McGilchrist (1994) best linear unbiased prediction (BLUP) based on the h-likelihood of Lee and Nelder (1996), or the Green (1987) penalized quasi-likelihood (PQL) [see also Breslow and Clayton (1993)]. Huber, Ronchetti and Victoria-Feser (2004) show that these estimators are all equal but differ from the LAMLE.

The results of Proposition 3 could, in principle, be used for inference when the sample sizes are large. When this is not that case, it is more suitable to use other techniques. For the correlation estimator $\hat{\rho}$, we propose to use the transformation function $\eta$ introduced by Fisher (1915) that stabilizes the variance of the estimator:

$$\eta(\rho) = \tanh^{-1}(\rho) = \frac{1}{2} \log\left(\frac{1+\rho}{1-\rho}\right),$$

and $\eta$ is approximately normal

$$\eta(\rho) \sim \mathcal{N}\left(\nu_\rho, \frac{1}{n-3}\right),$$

with $\nu_\rho = \tanh^{-1}(\rho) + \frac{\rho}{2(n-1)}$. A discussion about the Fisher transformation can be found in Efron (1982). In practice, we compute the variance of $\hat{\eta}$, which is simply $(n-3)^{-1}$, calculate its confidence interval, and transform it back to a confidence interval for $\hat{\rho}$.

For the other parameter estimators, we use a parametric bootstrap: first, we calculate the estimators from the observed sample and then, we generate 1000 new samples under the estimated distributions to get new estimators. We find the biases and endpoints of the confidence intervals using a bias-corrected acceleration ($BC_a$) technique as described in Efron (1987), Efron and Tibshirani (1993) and Shao and Tu (1995).

**6. Monte Carlo experiments.** In order to evaluate the performance of our model and our estimator in finite samples, we have performed a simulation study.[2] We consider the model (3) with $p = 10$, equal thresholds and parameter values given in Tables 1–3. With these two sets of parameters, we also choose different values for the correlation coefficient, namely, $\rho = -0.5, 0$ and $0.5$. The first set of parameters ($S1$) was chosen to match one of the real examples analyzed in Section 7 and the other ($S2$) to reflect what is sensible in practice, that is, a conservative attitude implying large probabilities associated to small or no changes. For each set of parameters, we simulated 500 samples of size $n = 30$, and computed the LAMLE of $\boldsymbol{\lambda}$ and the association measure $\rho$. The distribution of the sample bias estimates for the thresholds and loadings are presented for each estimator in the form of boxplots in Figures 3 and 4 (second latent variable) for a correlation $\rho = 0.5$ (for the other values of $\rho$ and the other loadings we find similar results). Figure 5 shows the boxplots for the estimated correlation $\hat{\rho}$ under the parameter set $S1$ (for the other set, results are similar). We can see that even for a relatively small sample size (given the size of the model), the performance is very good in that there is no apparent bias for all parameters, including $\rho$.

TABLE 1
*Thresholds for simulation $S1$*

| Thresholds | Cumulative probabilities |
|---|---|
| $-4.60$ | 0.01 |
| $-2.94$ | 0.05 |
| 0.85 | 0.70 |
| 4.60 | 0.99 |

TABLE 2
*Thresholds for simulation $S2$*

| Thresholds | Cumulative probabilities |
|---|---|
| $-2.19$ | 0.10 |
| $-1.39$ | 0.20 |
| 1.39 | 0.80 |
| 2.19 | 0.90 |

---

[2]The C code is provided as part of the supplemental material; see Huber, Scaillet and Victoria-Feser (2009b).

TABLE 3
*Loadings for simulation $S1$ and $S2$*

| Latent 1 | Latent 2 |
|---|---|
| 1.60 | 0.00 |
| 1.75 | 0.00 |
| 1.70 | 0.00 |
| 1.30 | 0.00 |
| 1.50 | 0.00 |
| 0.00 | 5.00 |
| 0.00 | 9.00 |
| 0.00 | 9.00 |
| 0.00 | 5.00 |
| 0.00 | 6.00 |

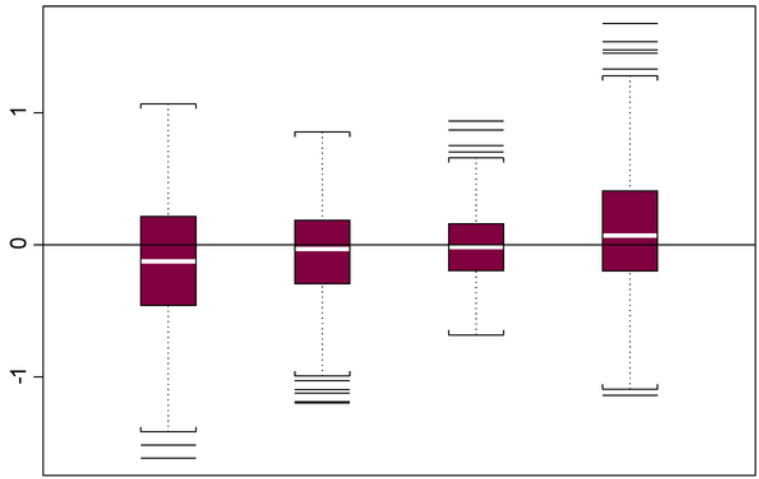

FIG. 3. *Distributions of threshold bias estimates for simulation $S1$ with $\rho = 0.5$.*

We have also studied the small sample performance of the probability coverage of 95% confidence intervals for $\hat{\rho}$ computed with the Fisher transformation, and have found a probability coverage of 84.9%.

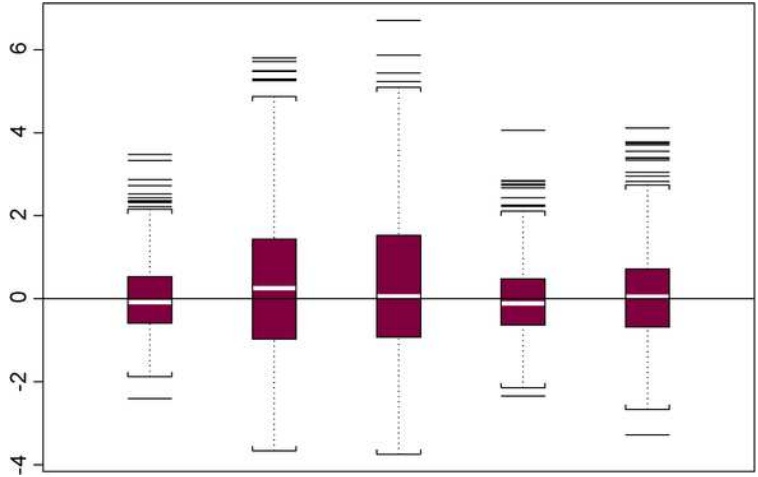

Fig. 4. *Distributions of loading estimates (second latent variable) for simulation $S1$ with $\rho = 0.5$.*

**7. Data analysis.** The estimated loadings for both broker–dealers are given in Tables 4 and 5 with biases and 95% confidence intervals, all com-puted via a parametric bootstrap. The estimated correlation between both latent variables for broker–dealers $A$ and $B$ is given in Table 6. The scores of the latent variables for both broker–dealers are displayed in Figures 6 and 7.

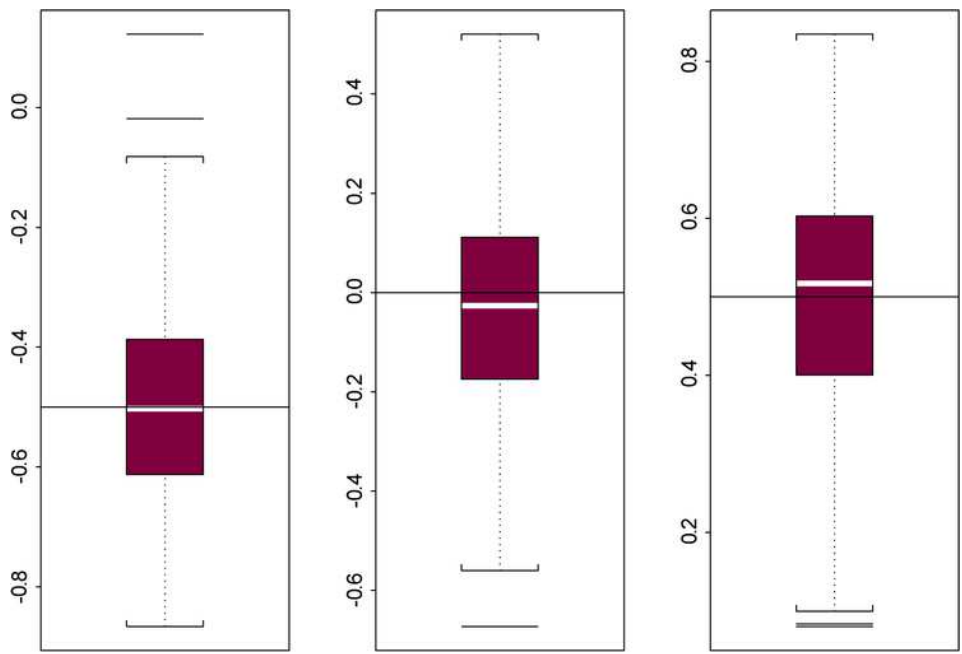

Fig. 5. *Distributions of correlation estimates for simulation $S1$ with, respectively, $\rho = -0.5$, $\rho = 0.0$ and $\rho = 0.5$.*

The study of the correlation estimates (see Table 6) indicates whether the broker–dealer forecasts match the actual market evolution. The correlation for both broker–dealers are significantly positive. We can therefore conclude that the forecasts are relatively accurate. Alternatively, we can look at the latent scores $\widehat{F}_{X_i}$ and $\widehat{F}_{Y_i}$ and see graphically how they evolve. For broker–dealer $A$, they are given in Figure 6 and for broker–dealer $B$ in Figure 7. For both broker–dealers, the evolution of the two lines (predicted and actual) is pretty similar, thus reflecting the fact that the predictions on the five stock markets and on the four bond markets are in phase with the actual evolutions of the latter. This reflects again the relatively accurate ability of the broker–dealers to predict the markets.

Tables 4 and 5 present the estimated loadings for broker–dealers $A$ and $B$, respectively. They give another type of information about the behavior of the broker–dealers. Indeed, the correlation reflects the ability of the broker–dealers to predict the changes in markets directions, but not necessarily the range of the changes. The latter can be inferred from the loadings because they act as a multiplicative factor of the latent variables. In other words, the latent variables give the directions of the market moves, while the loadings give the (average) ranges of these moves. In Tables 4 and 5 one can see that the loadings for the actual markets are systematically higher than the corresponding loadings related to the forecasts. This difference is certainly due to the forecasts being in general too conservative: although the direction of the movements are correctly predicted, their range is underestimated in all markets by the broker–dealers. We have already noticed this feature when analyzing the raw data in Figures 1 and 2.

TABLE 4
*Estimated loadings for broker–dealer A. The biases, lower ($l_{0.95}$) and upper ($u_{0.95}$) confidence bounds are computed with a parametric bootstrap*

| | **Predicted** | | | | **Observed** | | | |
|---|---|---|---|---|---|---|---|---|
| **Market** | **Estimator** | **Bias** | $l_{0.95}$ | $u_{0.95}$ | **Estimator** | **Bias** | $l_{0.95}$ | $u_{0.95}$ |
| CH | 1.596 | −0.078 | 0.441 | 3.071 | 5.557 | −0.280 | 3.716 | 10.069 |
| D | 1.762 | −0.063 | 0.456 | 3.197 | 8.925 | −1.091 | 5.399 | 17.079 |
| F | 1.725 | −0.066 | 0.411 | 3.407 | 8.982 | −1.102 | 5.493 | 17.709 |
| UK | 1.286 | −0.035 | 0.207 | 2.724 | 4.980 | −0.162 | 3.165 | 8.386 |
| USA | 1.506 | −0.103 | 0.302 | 3.023 | 6.222 | −0.442 | 4.178 | 11.628 |

TABLE 5
*Estimated loadings for broker–dealer B. The biases, lower ($l_{0.95}$) and upper ($u_{0.95}$) confidence bounds are computed with a parametric bootstrap*

| Market | Predicted | | | | Observed | | | |
|---|---|---|---|---|---|---|---|---|
| | Estimator | Bias | $l_{0.95}$ | $u_{0.95}$ | Estimator | Bias | $l_{0.95}$ | $u_{0.95}$ |
| CH | 0.632 | 0.038 | −0.985 | 1.824 | 2.971 | −0.082 | 1.836 | 5.608 |
| EU | 0.886 | −0.029 | −0.912 | 2.200 | 5.869 | −1.049 | 2.554 | 10.014 |
| UK | 0.544 | 0.029 | −0.930 | 1.732 | 5.419 | −0.897 | 2.662 | 9.234 |
| USA | 1.219 | −0.105 | −0.921 | 2.665 | 7.180 | −1.647 | 3.327 | 14.059 |

TABLE 6
*Estimated correlations between both latent variables*

| Broker–dealer A | | | | Broker–dealer B | | | |
|---|---|---|---|---|---|---|---|
| Estimator | Bias | $l_{0.95}$ | $u_{0.95}$ | Estimator | Bias | $l_{0.95}$ | $u_{0.95}$ |
| 0.398 | −0.034 | 0.013 | 0.722 | 0.320 | 0.069 | 0.027 | 0.691 |

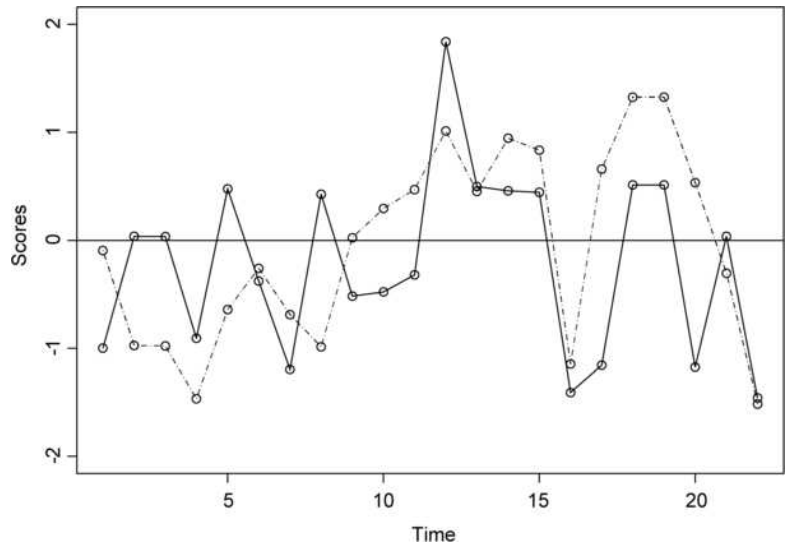

FIG. 6. *Estimated scores for broker–dealer A. The plain line is the forecast and the dotted line the actual level of the stock market.*

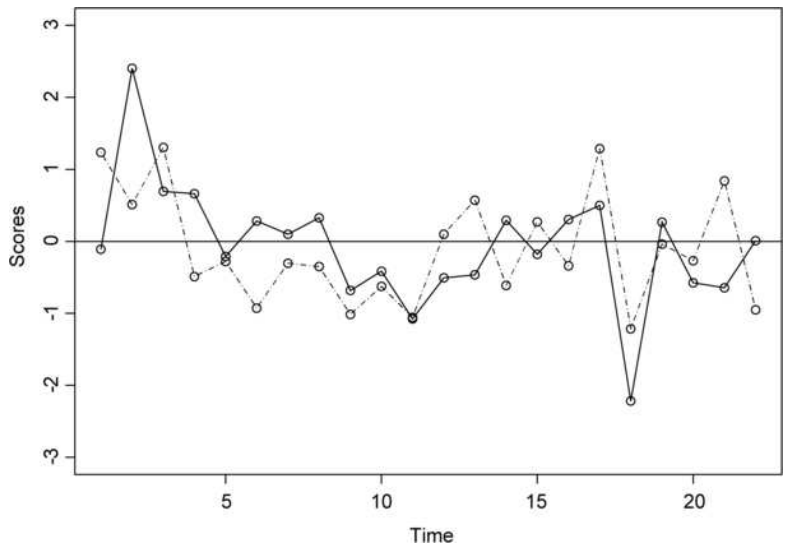

Fig. 7. *Estimated scores for broker–dealer B. The plain line is the forecast and the dotted line the actual level of the bond market.*

## SUPPLEMENTARY MATERIAL

**Supplement A: Datasets on the predictions by two broker–dealers and realized values on several markets** (DOI: 10.1214/08-AOAS213SUPPA; .zip). In this supplement, we provide a zip file containing two Excel files for the predictions and the realized market values of the two broker–dealers analyzed in this paper.

**Supplement B: C code for data analysis and simulations**
(DOI: 10.1214/08-AOAS213SUPPB; .zip). In this supplement we provide a zip file containing the source code in C for the programs used to analyze the datasets and to perform the simulation study in this paper.

**Supplement C: Technical developments and proofs**
(DOI: 10.1214/08-AOAS213SUPPC; .pdf). In this supplement we provide the technical developments for the likelihood comparison between the polychoric correlation and the GLLVM of Section 4, the development of the LAMLE for ordered multinomial distributed manifest variables as a complement of Section 5 and the proofs of Propositions 1–3.

## REFERENCES

AGRESTI, A. (1984). *Analysis of Ordinal Categorical Data*. Wiley, New York. MR0747468
AGRESTI, A. (1990). *Categorical Data Analysis*. Wiley, New York. MR1044993
AIGNER, D. J., HSIAO, C., KAPTEYN, A. and WANSBEEK, T. (1984). Latent variables models in econometrics. In *Handbook of Econometrics* **II** (Z. Griliches and M. Intriligator, eds.). North-Holland, Amsterdam. MR0772383

BARTHOLOMEW, D. J. and KNOTT, M. (1999). *Latent Variable Models and Factor Analysis. Kendall's Library of Statistics* **7**. Arnold, London. MR1711686

BLUME, M. (1993). Soft dollars and the brokerage industry. *Financial Analysts Journal* **49** 36–44.

BRESLOW, N. E. and CLAYTON, D. G. (1993). Approximate inference in generalized linear mixed models. *J. Amer. Statist. Assoc.* **88** 9–25.

COX, N. R. (1974). Estimation of the correlation between a continuous and a discrete variable. *Biometrics* **30** 171–178. MR0334376

DE BRUIJN, N. G. (1981). *Asymptotic Methods in Analysis*, 3rd ed. Dover, New York. MR0671583

EFRON, B. (1982). Transformation theory: How normal is a family of distributions? *Ann. Statist.* **10** 323–339. [Correction (1982) **10** 1032.] MR0653511

EFRON, B. (1987). Better bootstrap confidence intervals. *J. Amer. Statist. Assoc.* **82** 171–185. Comments on 186–200. MR0883345

EFRON, B. and TIBSHIRANI, R. (1993). *An Introduction to the Bootstrap*. Chapman & Hall, London. MR1270903

ELEFANT-YANNI, V., HUBER, P. and VICTORIA-FESER, M.-P. (2004). A simulation study to compare competing estimators in structural equation models with ordinal variables. Cahiers de recherche HEC no. 2004.10, Univ. Geneva.

EROSHEVA, E., FIENBERG, S. E. and JOUTARD, C. (2007). Describing disability through individual-level mixture models for multivariate binary data. *Ann. Appl. Statist.* **1** 502–537.

FIELD, C. A. and RONCHETTI, E. (1990). *Small Sample Asymptotics*. IMS, Haywood, CA. MR1088480

FIELDING, A. and YANG, M. (2005). Generalized linear mixed models for ordered responses in complex multilevel structures: Effects beneath the school or college in education. *J. Roy. Statist. Soc. Ser. A* **168** 159–183. MR2113233

FISHER, R. A. (1915). Frequency distribution of the values of the correlation coefficient in samples from an indefinitely large population. *Biometrika* **10** 507–521.

GOODMAN, L. A. and KRUSKAL, W. H. (1959). Measures of association for cross classifications. *J. Amer. Statist. Assoc.* **54** 123–163.

GOODMAN, L. A. and KRUSKAL, W. H. (1979). *Measures of Association for Cross Classifications*. Springer, New York. MR0553108

GREEN, P. J. (1987). Penalized likelihood for general semi-parametric regression models. *Internat. Statist. Rev.* **55** 245–259. MR0963142

HUBER, P., RONCHETTI, E. and VICTORIA-FESER, M.-P. (2004). Estimation of generalized latent trait models. *J. Roy. Statist. Soc. Ser. B* **66** 893–908. MR2102471

HUBER, P., SCAILLET, O. and VICTORIA-FESER, M.-P. (2009a). Supplement to "Assessing multivariate predictors of financial market movements: A latent factor framework for ordinal data." DOI: 10.1214/08-AOAS213SUPPA.

HUBER, P., SCAILLET, O. and VICTORIA-FESER, M.-P. (2009b). Supplement to "Assessing multivariate predictors of financial market movements: A latent factor framework for ordinal data." DOI: 10.1214/08-AOAS213SUPPB.

HUBER, P., SCAILLET, O. and VICTORIA-FESER, M.-P. (2009c). Supplement to "Assessing multivariate predictors of financial market movements: A latent factor framework for ordinal data." DOI: 10.1214/08-AOAS213SUPPC.

JOHNSEN, B. (1994). Property rights to investment research: The agency costs of soft dollar brokerage. *The Yale Journal of Regulations* **11** 57–73.

JÖRESKOG, K. G. (1994). On the estimation of polychoric correlations and their asymptotic covariance matrix. *Psychometrika* **59** 381–389. MR1309413

JÖRESKOG, K. G. and MOUSTAKI, I. (2001). Factor analysis for ordinal variables: A comparison of three approaches. *Multivariate Behavioral Research* **36** 347–387.

KELLER, W. J. and WANSBEEK, T. (1983). Multivariate methods for quantitative and qualitative data. *J. Econometrics* **22** 91–111. MR0719129

KENDALL, M. G. (1945). The treatment of ties in rank problems. *Biometrika* **33** 239–251. MR0016601

LEE, S.-Y. and POON, W.-Y. (1986). Maximum likelihood estimation of polychoric correlations. *Psychometrika* **51** 113–121. MR0829891

LEE, Y. and NELDER, J. A. (1996). Hierarchical generalized linear models. *J. Roy. Statist. Soc. Ser. B* **58** 619–678. MR1410182

LORD, F. M. and NOVICK, M. E. (1968). *Statistical Theories of Mental Test Scores*. Addison-Wesley, Reading, MA.

MARDIA, K. V., KENT, J. T. and BIBBY, J. M. (1979). *Multivariate Analysis*. Academic Press, London. MR0560319

MCCULLAGH, P. and NELDER, J. A. (1989). *Generalized Linear Models*, 2nd ed. Chapman & Hall, London.

MCFADDEN, D. (1984). Econometric analysis of qualitative response models. In *Handbook of Econometrics* **II** (Z. Griliches and M. Intriligator, eds.). North-Holland, Amsterdam. MR772383

MCGILCHRIST, C. A. (1994). Estimation in generalized mixed models. *J. Roy. Statist. Soc. Ser. B* **56** 61–69. MR1257795

MOUSTAKI, I. (1996). A latent trait and a latent class model for mixed observed variables. *British J. Math. Statist. Psych.* **49** 313–334.

MOUSTAKI, I. and KNOTT, M. (2000). Generalized latent trait models. *Psychometrika* **65** 391–411. MR1792703

MUTHÉN, B. (1984). A general structural equation model with dichotomous, ordered categorical and continuous latent variables indicators. *Psychometrika* **49** 115–132.

NEWEY, W. and MCFADDEN, D. (1994). Large sample estimation and hypothesis testing. In *Handbook of Econometrics* **IV** (R. F. Engle and D. McFadden, eds.). North-Holland, Amsterdam. MR1315971

OLSSON, U. (1979). Maximum likelihood estimation of the polychoric correlation coefficient. *Psychometrika* **44** 443–460. MR0554892

POON, W.-Y. and LEE, S.-Y. (1987). Maximum likelihood estimation of multivariate polyserial and polychoric correlation coefficients. *Psychometrika* **52** 409–430. [Correction (1988) **53** 301.] MR0914463

RABE-HESKETH, S., SKRONDAL, A. and PICKLES, A. (2002). Reliable estimation of generalized linear mixed models using adaptive quadrature. *The Stata Journal* **2** 1–21.

SECURITIES AND EXCHANGE COMMISSION (1998). Inspection report on the soft dollar practices of broker–dealers, investment advisers and mutual funds.

SHAO, J. and TU, D. (1995). *The Jackknife and Bootstrap*. Springer, New York. MR1351010

SKRONDAL, A. and RABE-HESKETH, S. (2004). *Generalized Latent Variable Modeling: Multilevel, Longitudinal, and Structural Equation Models*. Chapman & Hall, London. MR2059021

TATE, R. F. (1955a). Applications of correlation models for biserial data. *J. Amer. Statist. Assoc.* **50** 1078–1095.

TATE, R. F. (1955b). The theory of correlation between two continuous variable when one is dichotomized. *Biometrika* **42** 205–216. MR0070916

TIERNEY, L. and KADANE, J. B. (1986). Accurate approximations for posterior moments and marginal densities. *J. Amer. Statist. Assoc.* **81** 82–86. MR0830567

Wu, H., Yuan, M., Kaech, S. M. and Halloran, M. E. (2007). A statistical analysis of memory cd8 t cell differentiation: An application of a hierarchical state space model to a short time course microarray experiment. *Ann. Appl. Statist.* **2** 442–458.

Zaslavsky, A. M. (2007). Using hierarchical models to attribute sources of variation in consumer assessments of health care. *Statistics in Medicine* **26** 1885–1900. MR2359199

Zhou, Q. and Wong, W. H. (2007). Coupling hidden Markov models for the discovery of cis-regulatory modules in multiple species. *Ann. Appl. Statist.* **1** 36–65. MR2393840

P. Huber
Cinetics SA
Route du Vélodrome 66
1228 Plan-les-Ouates
Switzerland
E-mail: ph_geneve@hotmail.com

O. Scaillet
HEC
University of Geneva
and
Swiss Finance Institute
Bd du Pont d'Arve 40
CH-1211 Geneve 4
Switzerland
E-mail: olivier.scaillet@unige.ch

M.-P. Victoria-Feser
HEC
University of Geneva
Bd du Pont d'Arve 40
CH-1211 Geneve 4
Switzerland
E-mail: maria-pia.victoriafeser@unige.ch